\documentclass{article}

\usepackage[dvips]{graphicx}  

\begin{document}

\newcommand{\rum}{\rule{0.5pt}{0pt}}

\newcommand{\rub}{\rule{1pt}{0pt}}

\newcommand{\rim}{\rule{0.3pt}{0pt}}

\newcommand{\numtimes}{\mbox{\raisebox{1.5pt}{${\scriptscriptstyle \times}$}}}

\renewcommand{\refname}{References}

\begin{center}

{\Large\bf 3-Space In-Flow Theory of Gravity: Boreholes, Blackholes and the Fine Structure Constant
\rule{0pt}{13pt}}\par

\bigskip

Reginald T. Cahill \\ 

{\small\it School of Chemistry, Physics and Earth Sciences, Flinders University,

Adelaide 5001, Australia\rule{0pt}{13pt}}\\

\raisebox{-1pt}{\footnotesize E-mail: Reg.Cahill@flinders.edu.aul}\par

\bigskip\smallskip

{\small\parbox{11cm}{%
A  theory of 3-space explains the phenomenon of gravity as arising from the time-dependence and
inhomogeneity of the differential flow of this 3-space. The emergent theory of gravity has two
gravitational constants:
$G_N$ - Newton's constant, and a dimensionless constant $\alpha$. Various experiments and astronomical
observations have shown that
$\alpha$ is the fine structure constant $\approx 1/137$. Here we analyse the Greenland Ice Shelf and
Nevada Test Site borehole $g$ anomalies, and confirm with increased precision this value of $\alpha$. 
This and other successful tests of this theory of gravity, including  the
supermassive black holes in globular clusters and galaxies, and the `dark-matter' effect in
spiral galaxies, demonstrates the validity of this theory of gravity. This success implies that 
 Newtonian gravity was fundamentally flawed from the beginning. 

\rule[0pt]{0pt}{0pt}}}\bigskip

\end{center}

\section{Introduction\label{sect:introduction}}
In the Newtonian theory of gravity \cite{Newton} the Newtonian gravitational constant $G_N$ determining the strength
of this phenomenon is difficult to measure because of the extreme weakness of gravity. Originally determined in
laboratory experiments by Cavendish  \cite{Cavendish} in 1798 using a torsion balance, Airy \cite{Airy} in
1865 presented a different method which compared the gravity gradients above and below the surface of the
earth.  Then if the matter density within the neighbourhood of the measurements is sufficiently uniform, or
 at most is horizontally layered and known, then  such measurements then permitted $G_N$ to be
determined, as discussed below, if Newtonian gravity was indeed correct.  Then the mass of the earth can be
computed from the value of $g$ at the earth's surface.  However two anomalies have emerged for these two
methods: (i) the  Airy method has given gravity gradients that are inconsistent with Newtonian gravity, and
 (ii) the laboratory measurements of $G_N$ using various geometries for the test masses have not converged
despite ever increasing experimental sophistication and precision.  There are other anomalies involving
gravity such as the so-called `dark-matter' effect in spiral galaxies, the systematic effects related to the
supermassive blackholes in globular clusters and elliptical galaxies, the Pioneer 10/11 deceleration
anomaly, the so-called  galactic `dark-matter' networks, and others, all suggest that the phenomenon of
gravity has not been understood even in the non-relativistic regime, and that a significant dynamical process
has been overlooked in the Newtonian theory of gravity, and which is also missing from General Relativity.  

The discovery of this missing dynamical process arose from  experimental evidence  \cite{Book,MM,AMGE} that
 a complex dynamical 3-space underlies reality.  The evidence involves the repeated detection
of the  motion of the earth relative to that 3-space using Michelson interferometers  operating in gas
mode \cite{MM}, particularly the experiment by Miller   in 1925/26 at Mt.Wilson, and the
coaxial cable  RF travel time measurements by Torr and Kolen in Utah in 1985, and the DeWitte experiment
in 1991 in Brussels \cite{MM}. In  all  7 such experiments are consistent with respect to speed and
direction. It has been shown that effects caused by motion relative to this 3-space can mimic the
formalism of spacetime, but that it is the 3-space that is `real', simply because it is directly
observable \cite{Book}. 

The 3-space is in differential motion, that is one part has a velocity relative
to  other parts, and so involves a velocity field ${\bf v}({\bf r},t)$ description. To be
specific this velocity field must be described relative to a frame of observers, but the
formalism is such that the dynamical equations for this velocity field must transform
covariantly under a change of observer. 
It has been shown   \cite{Book,DMtrends} that the phenomenon of gravity is a consequence of the 
time-dependence
and inhomogeneities of  ${\bf v}({\bf r},t)$.  So the dynamical equations for  ${\bf v}({\bf
r},t)$ give rise to a new theory of gravity when combined with the generalised Schr\"{o}dinger equation, and
the generalised Maxwell and Dirac equations \cite{Schrod}. The equations for  ${\bf v}({\bf
r},t)$ involve the  gravitational constant\footnote{This is different from the Newtonian effective
gravitational constant $G_N$ defined later.}
$G$ and a dimensionless constant that determines the strength of a new 3-space self-interaction effect, which is
missing from both Newtonian Gravity and General Relativity. Experimental data has revealed
\cite{Book,alpha,DMtrends}  the remarkable discovery that this constant is the fine structure constant $\alpha
\approx e^2/\hbar c \approx 1/137$.  This dynamics then explains numerous gravitational anomalies, such as the
borehole $g$ anomaly, the so-called `dark matter' anomaly in the rotation speeds of spiral galaxies, and that the
effective mass of the necessary black holes at the centre of spherical matter systems, such as globular clusters and
spherical galaxies, is $\alpha/2$ times the total mass of these systems. This prediction has been confirmed by
astronomical  observations
\cite{BH}.  

Here we analyse the Greenland  and
Nevada Test Site borehole $g$ anomalies, and confirm with increased precision this value of $\alpha$. 

The occurrence of $\alpha$ suggests that space is itself a quantum system undergoing on-going
classicalisation.  Just such a proposal has arisen in {\it Process Physics} \cite{Book} which is
an information-theoretic modelling of reality. There quantum space and matter arise in terms of
the Quantum Homotopic Field Theory (QHFT)  which, in turn, may be related to the standard model of
matter. In the QHFT space at this quantum level is best described as a `quantum foam'. So 
 we  interpret the observed fractal\footnote{The fractal property of 3-space was established from the DeWitte data
in \cite{Schrod}.} 3-space as a classical approximation to this `quantum foam' \cite{Schrod}.

\section{Dynamical 3-Space\label{sect:space}}

Relative to some observer 3-space is described by a velocity field  ${\bf v}({\bf r},t)$.
It is important to note that the coordinate ${\bf r}$  is not itself 3-space, rather it
is merely a label for an element of 3-space that has  velocity ${\bf v}$, relative to some
observer.  Also it is important to appreciate that this `moving' 3-space is not
itself  embedded in a `space'; the 3-space is all there is, although as noted above its deeper
structure is that of a `quantum foam'.  

\begin{figure}
\hspace{45mm}\includegraphics[scale=1.1]{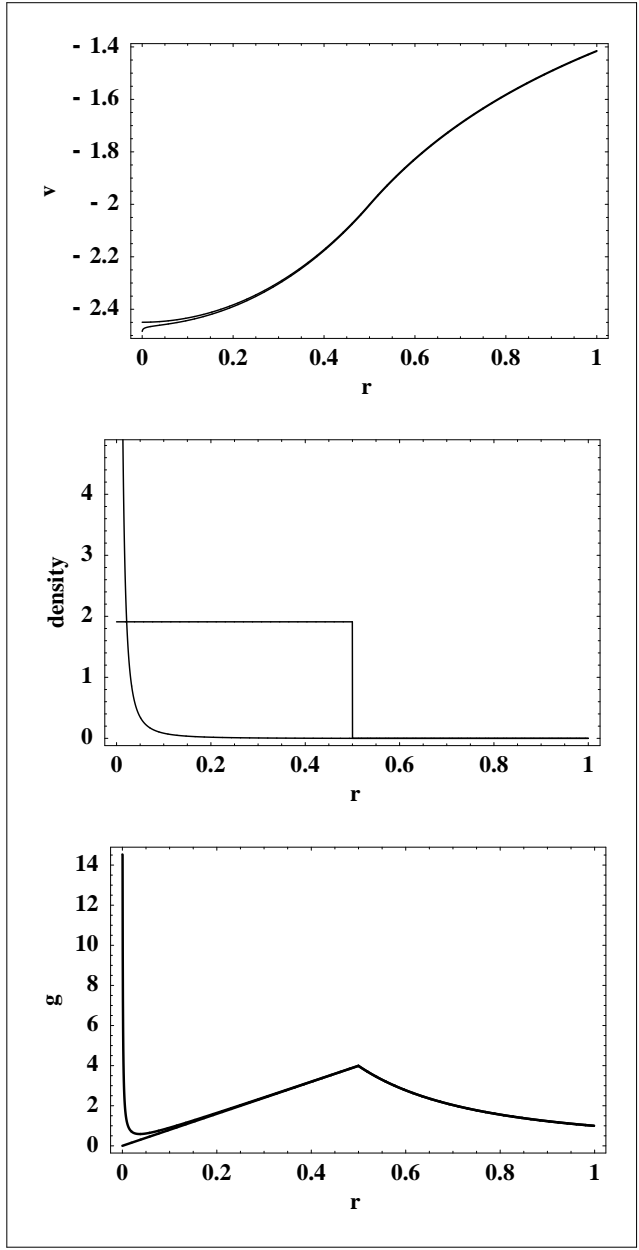}
\caption{\small{ Upper plot shows speeds from numerical iterative  solution of (\ref{eqn:E7}) for a solid sphere with
uniform density and radius $r=1$ for (i) upper curve the case $\alpha=0$ corresponding to Newtonian gravity, and (ii)
lower curve with $\alpha=1/137$. These solutions ony differ significantly near $r=0$. Middle  plot shows matter density
and `dark matter' density $\rho_{DM}$, from (\ref{eqn:E5}), with arbitrary scales.  Lower plot shows the acceleration
from (\ref{eqn:E3}) for (i) the Newtonian in-flow  from the upper plot, and (ii) from the $\alpha=1/137$ case. The
difference is only significant near $r=0$. The accelerations begin to differ just inside the surface of the sphere at
$r=1$, according to (\ref{eqn:E15}).  This difference is the origin of the borehole $g$ anomaly, and permits the
determination of the value of $\alpha$ from observational data. This generic singular-$g$ behaviour, at $r=0$,
is seen in the earth, in globular clusters and in galaxies. }
\label{fig:earthplots}}\end{figure}

In the case of zero vorticity $\nabla\times{\bf v}={\bf 0}$ the 3-space dynamics is given by \cite{Book,DMtrends}, in
the non-relativistic limit,
\begin{equation}
\nabla.\left(\frac{\partial {\bf v} }{\partial t}+({\bf v}.{\bf \nabla}){\bf v}\right)
+\frac{\alpha}{8}\left((tr D)^2 - tr(D^2)\right)=
-4\pi G\rho,
\label{eqn:E1}\end{equation}
where $\rho$ is the matter density, and where 
\begin{equation} D_{ij}=\frac{1}{2}\left(\frac{\partial v_i}{\partial x_j}+
\frac{\partial v_j}{\partial x_i}\right).
\label{eqn:E2}\end{equation}
The acceleration of an element of space is given by the Euler form
\begin{eqnarray}
{\bf g}({\bf r},t)&\equiv&\lim_{\Delta t \rightarrow 0}\frac{{\bf v}({\bf r}+{\bf v}({\bf r},t)\Delta t,t+\Delta
t)-{\bf v}({\bf r},t)}{\Delta t} \nonumber \\
&=&\frac{\partial {\bf v}}{\partial t}+({\bf v}.\nabla ){\bf v}
\label{eqn:E3}\end{eqnarray} 
\begin{figure*}[t]
\hspace{15mm}\includegraphics[scale=0.44]{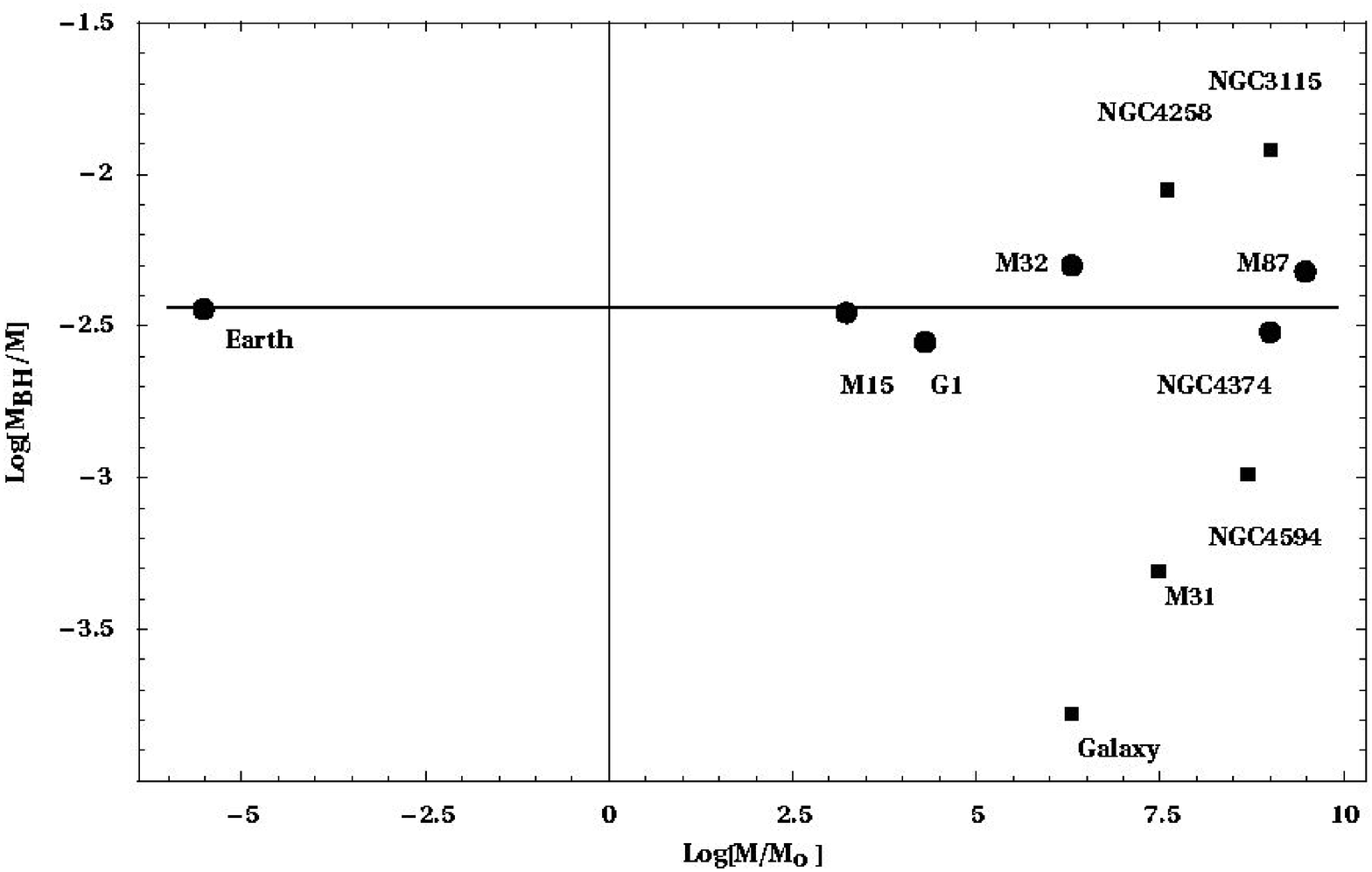}
\caption{\small{The data shows $\mbox{Log}_{10}[M_{BH}/M]$ for the `blackhole' or `dark matter' masses $M_{BH}$  for
a variety of spherical matter systems with masses $M$, shown by solid circles, plotted against 
$\mbox{Log}_{10}[M/M_0]$, where
$M_0$ is the solar mass, showing agreement with the `$\alpha/2$-line' ($\mbox{Log}_{10}[\alpha/2]=-2.44$) predicted by
(\ref{eqn:E10}), and ranging over 15 orders of magnitude. The `blackhole' effect is the same phenomenon as the
`dark matter' effect. The data ranges from the earth, as observed by the bore hole $g$ anomaly, to globular cluster
M15  and G1, and then to spherical `elliptical' galaxies M32 (E2), NGC
4374 (E1)  and M87 (E0). Best fit to the data from these star systems gives $\alpha=1/134$, while for the earth data
in Figs.\ref{fig:Greenland}, \ref{fig:Nevada1}, \ref{fig:Nevada2}  give
$\alpha=1/137$. In these systems the `dark
matter' or `black hole' spatial self-interaction effect is induced by the matter. For the spiral galaxies, shown by
the filled boxes, where here $M$ is  the bulge mass, the blackhole masses do not correlate with the
`$\alpha/2$-line'. This is because these systems form by matter in-falling to a primordial blackhole, and so these
systems are more contingent. For spiral galaxies this dynamical effect manifests most clearly via the non-Keplerian
rotation-velocity curve, which decrease asymptotically very slowly.  See \cite{BH} for references to the
data.}
\label{fig:blackholes}}\end{figure*}
It was shown in \cite{Schrod}  that matter has the same
acceleration\footnote{Except for the acceleration component induced by  vorticity.} as (\ref{eqn:E3}), which gave
a derivation of the equivalence principle as a quantum effect in the Schr\"{o}dinger equation when uniquely
generalised to include the interaction of the quantum system with the 3-space.  These forms are mandated by Galilean
covariance under change of observer\footnote{However this does not exclude  so-called relativistic effects, such as
the length contraction of moving rods or the time dilations of moving clocks.}.   This minimalist non-relativistic
modelling of the dynamics for the velocity field gives a direct account of the various phenomena noted above. A
generalisation to include vorticity and relativistic effects of the motion of matter through this 3-space is given in
\cite{Book}. From (\ref{eqn:E1}) and (\ref{eqn:E3}) we obtain that
\begin{equation}
\nabla.{\bf g}=-4\pi G\rho-4\pi G \rho_{DM},
\label{eqn:E4}\end{equation}
where 
\begin{equation}
\rho_{DM}({\bf r})=\frac{\alpha}{32\pi G}( (tr D)^2-tr(D^2)).  
\label{eqn:E5}\end{equation}
In this form we see that if $\alpha\rightarrow 0$, then the acceleration of the 3-space elements is given
by Newton's {\it Universal Law of Gravitation}, in differential form. But for a non-zero $\alpha$ we see that the
3-space acceleration has an additional effect, from the $\rho_{DM}$ term, which is an effective `matter density' that
mimics the new self-interaction dynamics.  This has been shown to be the origin of the so-called `dark
matter' effect in spiral galaxies.  It is important to note that (\ref{eqn:E4})  does not determine ${\bf
g}$ directly; rather the velocity dynamics in (\ref{eqn:E1}) must be solved, and then with ${\bf g}$
subsequently determined from (\ref{eqn:E3}).  Eqn.(\ref{eqn:E4}) merely indicates that the resultant
non-Newtonian ${\bf g}$  could be mistaken as the result of a new form of matter, whose density is given
by $\rho_{DM}$. Of course the saga of `dark matter' shows that this actually happened, and that there has
been a misguided and fruitless search for such `matter'.

\section{Airy Method for Determining  $\alpha$}

We now show that the Airy method actually gives a technique for determining the value of $\alpha$ from earth based
borehole gravity measurements. For a time-independent velocity field (\ref{eqn:E1}) may be written in the integral
form
\begin{equation}\label{eqn:E6}
|{\bf v}({\bf r})|^2=2G\int d^3
r^\prime\frac{\rho({\bf r}^\prime)+\rho_{DM}({\bf r}^\prime)}{|{\bf r}-{\bf r}^\prime|}.
\end{equation}
When the matter density of the earth is assumed to be spherically symmetric, and that the velocity field is
now radial\footnote{This in-flow is additional to the observed velocity of the earth through 3-space.}  
(\ref{eqn:E6}) becomes
\begin{equation}
v(r)^2=\frac{8\pi G}{r}\int_0^r s^2 \left[\rho(s)+\rho_{DM}(s)\right]ds +8\pi G\int_r^\infty s
\left[\rho(s)+\rho_{DM}(s)\right]ds, 
\label{eqn:E7}\end{equation}
 where, with $v^\prime=dv(r)/dr$,
\begin{equation}
\rho_{DM}(r)= \frac{\alpha}{32\pi G}\left(\frac{v^2}{2r^2}+ \frac{vv^\prime}{r}\right).
\label{eqn:E8}\end{equation}
Iterating (\ref{eqn:E7}) once we find to 1st order in $\alpha$ that 
\begin{equation}
\rho_{DM}(r)=\frac{\alpha}{2r^2}\int_r^\infty s\rho(s)ds+O(\alpha^2),
\label{eqn:E9}\end{equation}
so that in spherical systems the `dark matter' effect is concentrated near the centre, and we find that the total
`dark matter' is
\begin{eqnarray}
M_{DM} &\equiv& 4\pi\int_0^\infty r^2\rho_{DM}(r)dr  \nonumber \\ &=&\frac{4\pi\alpha}{2}\int_0^\infty
r^2\rho(r)dr+O(\alpha^2) \nonumber \\ &=&\frac{\alpha}{2}M+O(\alpha^2), 
\label{eqn:E10}\end{eqnarray}
where $M$ is the total amount of (actual) matter. Hence to $O(\alpha)$   $M_{DM}/M=\alpha/2$ independently of the matter
density profile. This turns out to be  a very useful property as complete knowledge of the density profile is then
not required in order to analyse observational data. As seen in Fig.\ref{fig:earthplots} the singular behaviour of
both $v$ and $g$ means that there is a {\it blackhole}\footnote{These are called {\it blackholes} because there is
an event horizon, but in all other aspects differ from the {\it blackholes} of General Relativity.} singularity at
$r=0$. Interpreting  $M_{DM}$ in (\ref{eqn:E10}) as the mass of the blackholes observed in the globular clusters
$M15$ and $G1$ and in the highly spherical `elliptical' galaxies M32, M87 and NGC 4374, we
obtained \cite{BH} $\alpha \approx 1/134$, as shown in Fig.\ref{fig:blackholes}.

From (\ref{eqn:E3}), which is also the acceleration of matter \cite{Schrod}, the gravity acceleration\footnote{We
now  use the convention that $g(r)$ is positive if it is radially inward.} is found to be, to 1st order in
$\alpha$,  and using that $\rho(r)=0$ for $r>R$, where $R$ is the radius of the earth,
\begin{equation}
g(r)=\left\{ \begin{tabular}{ l} 
$\displaystyle{\frac{(1+\displaystyle{\frac{\alpha}{2}}) GM}{r^2}, \mbox{\ \ } r > R,}$  \\  
$\displaystyle{\frac{4\pi G}{r^2}\int_0^rs^2\rho(s)ds +\frac{2\pi\alpha G}{r^2}\int_0^r\left(\int_s^R s^\prime \rho(s^\prime)ds^\prime
\right) ds},\mbox{\  } r< R$.
\end{tabular}\right.   
\label{eqn:E11}\end{equation}
This gives   Newton's `inverse square law' for $r > R$,  even when $\alpha \neq 0$, which explains why the
3-space self-interaction dynamics did not overtly manifest in the analysis of planetary orbits by Kepler and then
Newton. However inside the earth (\ref{eqn:E11}) shows that  $g(r)$ differs from the Newtonian theory,
corresponding to $\alpha=0$,  as in Fig\ref{fig:earthplots}, and it is this effect that allows the determination of
the value of  $\alpha$ from the Airy method.

Expanding  (\ref{eqn:E11}) in $r$ about the surface, $r=R$, we obtain,  to 1st order in
$\alpha$ and  for an arbitrary density profile, but not any retaining density gradients at the surface,
\begin{equation}
g(r)=\!\left\{ \begin{tabular}{ l} 
$\!\!\!\!\! \displaystyle{\frac{G_N M}{R^2}-\frac{2 G_N M}{R^3}(r-R),
\mbox{\ \ \ \  \ \ \ \ \  } r > R,}$ 
\\  \\ 
$\!\!\!\!\! \displaystyle{\frac{G_N M}{R^2}-\left(\frac{2 G_N M}{R^3}-4\pi(1-\frac{\alpha}{2})G_N\rho\right)\!(r-R),\mbox{\ \ \ } r<
R}$
\end{tabular}\right.   
\label{eqn:E12}\end{equation}
where  $\rho$ is the matter density at the surface, $M$ is the total matter mass of the earth, and where we have
defined
\begin{equation}G_N\equiv(1+\frac{\alpha}{2})G.\label{eqn:E13}\end{equation}
The corresponding Newtonian gravity expression is obtained by taking the limit $\alpha\rightarrow 0$,
\begin{equation}
g_N(r)=\!\left\{ \begin{tabular}{ l} 
$\!\!\!\!\!\displaystyle{\frac{G_N M}{R^2}-\frac{2 G_N M}{R^3}(r-R), \mbox{\ \ \ \  } r > R,}$  \\  \\
$\!\!\!\!\! \displaystyle{\frac{G_N M}{R^2}-\left(\frac{2 G_N M}{R^3}-4\pi G_N\rho\right)\!(r-R),\mbox{\ \  }r< R}$
\end{tabular}\right.   
\label{eqn:E14}\end{equation}
Assuming Newtonian gravity (\ref{eqn:E14})
then means that from  the measurement of difference between  the above-ground and  below-ground gravity gradients,
namely $4\pi G_N\rho$,  and also measurement of the matter density, permit the determination of  $G_N$.
This is the basis of the Airy method for determining $G_N$ \cite{Airy}.
\begin{figure}[t]
\hspace{45mm}\includegraphics[scale=0.27]{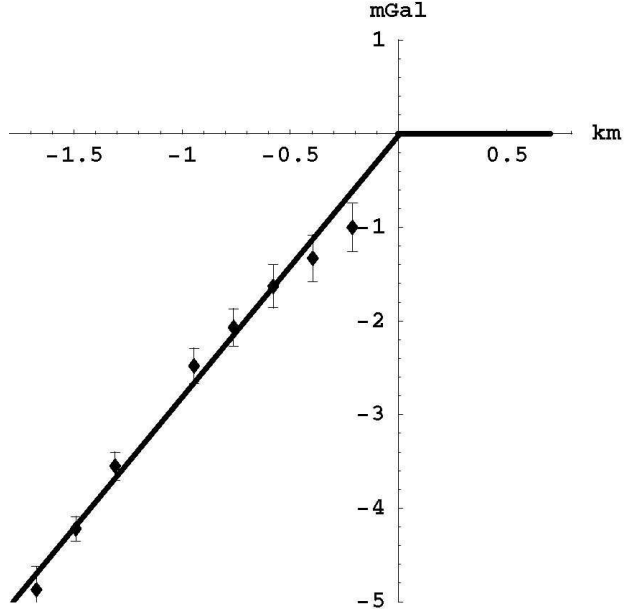}
\caption{\small{ The data shows the gravity residuals for the Greenland Ice Shelf \cite{Ander89} Airy measurements of
the
$g(r)$  profile,  defined as $\Delta g(r) = g_{Newton}-g_{observed}$, and measured in mGal (1mGal $ =10^{-3}$ cm/s$^2$)
and   plotted against depth in km. The gravity residuals have been offset. The borehole effect is that Newtonian
gravity and the new theory differ only beneath the surface, provided that the measured above surface gravity gradient 
is used in  both theories.  This then gives the horizontal line above the surface. Using (\ref{eqn:E15}) we obtain
$\alpha^{-1}=137.9 \pm  5$ from fitting the slope of the data, as shown. The non-linearity  in the data arises from
modelling corrections for the gravity effects of the   irregular sub ice-shelf rock  topography.}
\label{fig:Greenland}}\end{figure}

When analysing the borehole data it has been  found \cite{Ander89,Thomas90} that  the observed difference of the
gravity  gradients was inconsistent with $4\pi G_N\rho$ in  (\ref{eqn:E14}), in that it was not given by the
laboratory value of $G_N$ and the measured matter density.  This is known as the borehole
$g$ anomaly and which attracted much interest in the 1980's. 
The borehole data papers  \cite{Ander89,Thomas90} report the discrepancy, i.e. the anomaly or the gravity residual as it
is called, between the Newtonian prediction and the measured below-earth gravity gradient. Taking the difference
between 
 (\ref{eqn:E12})  and (\ref{eqn:E14}), assuming the same unknown value of $G_N$ in both,  we obtain an expression for
the gravity residual
\begin{equation}
\Delta g(r)\equiv  g_N(r)-g(r)=\left\{ \begin{tabular}{ l} 
$\mbox{\ \ }0, \mbox{\ \ \ \ \ \ \ } r> R,$  \\   
$2\pi\alpha G_N\rho(r-R), \mbox{\  } r < R.$
\end{tabular}\right. 
\label{eqn:E15}\end{equation}

When $\alpha\neq 0$ we have a two-parameter theory of gravity, and from (\ref{eqn:E12}) we see that measurement
of the difference between the above ground and below ground gravity gradients is $4\pi(1-\frac{\alpha}{2}) G_N\rho$, and
this  is not sufficient to determine both $G_N$ and $\alpha$, given $\rho$, and so the Airy method is now understood
not to be a complete measurement by itself, i.e. we need to combine it with other measurements.  If we now use
laboratory Cavendish experiments to determine $G_N$, then from the borehole  gravity residuals we can determine the
value of
$\alpha$, as already indicated in \cite{alpha,DMtrends}.  As discussed in Sect.\ref{sect:Gdata}  these Cavendish
experiments can only determine $G_N$ up to corrections of order $\alpha/4$, simply because the analysis of the data from
these experiments assumed the validity of Newtonian gravity. So the analysis of the borehole residuals will give the
value of $\alpha$ up to $O(\alpha^2)$ corrections, which is consistent with the $O(\alpha)$ analysis reported above.

\begin{figure}[t]
\hspace{45mm}\includegraphics[scale=0.27]{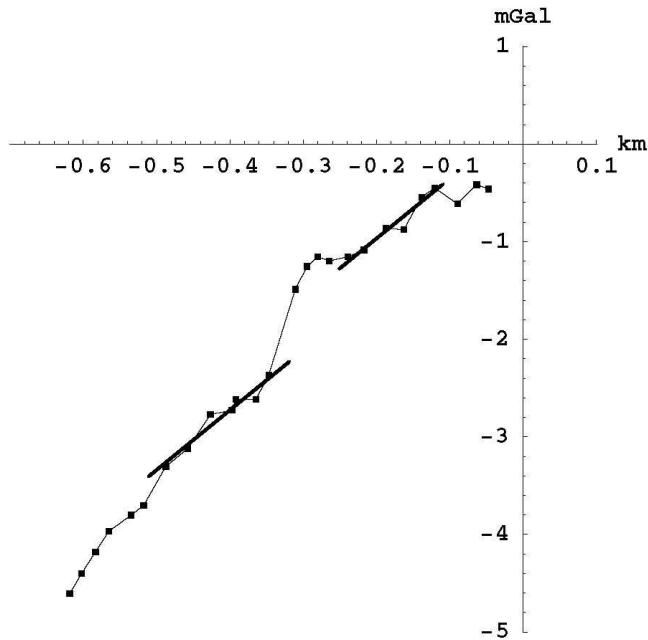}
\caption{\small{ The data shows the gravity residuals for the Nevada U20AK borehole  Airy measurements of the
$g(r)$  profile \cite{Thomas90},  defined as
$\Delta g(r) = g_{Newton}-g_{observed}$, and measured in mGal, plotted against depth in km.
This residual shows regions of linearity interspersed with regions of non-linearity, presumably arising from layers with
a density different from the main density of 2000 kg/m$^3$. Density changes generate a change in the (arbitrary)
residual offset.  From a least-squares simultaneous fit of the linear form in (\ref{eqn:E15}) to the slope of the
 four linear regions  in this data and that in Fig.\ref{fig:Nevada2} for the data from borehole U20AL, we obtain
$\alpha^{-1}=136.8\pm 3$. The two fitted regions of data are shown by the two straight lines here and in 
Fig.{\ref{fig:Nevada2}.}}
\label{fig:Nevada1}}\end{figure} 

\section{Greenland Ice Shelf Borehole Data}

Gravity  residuals  from a bore hole
into the Greenland Ice Shelf  were determined  down to a depth of 1.5 km by Ander {\it et al.} \cite{Ander89} in
1989. The observations were made at the Dye 3 2033 m deep borehole, which reached the basement rock. This borehole is
60 km south of the Arctic Circle and 125 km inland from the Greenland east coast at an elevation of 2530 m. It was
believed that the ice  provided an opportunity to use the Airy method to determine $G_N$, but now it is
understood that in fact the borehole residuals permit the determination of $\alpha$, given a laboratory value for
$G_N$. Various  steps were taken to remove unwanted effects, such as imperfect knowledge of the ice density and,
most dominantly, the terrain effects which arises from ignorance of the profile and density inhomogeneities of the
underlying rock.  The borehole gravity meter  was calibrated by comparison with
 an absolute gravity meter. The ice density depends on pressure, temperature and air content, with the density rising to
its average value of $\rho=920$ kg/m$^3$ within some 200 m  of the surface, due to compression of the trapped air
bubbles. This surface gradient in the density has been modelled by the author, and is not large enough the affect the
results. The leading source  of uncertainty was from the gravitational effect of the bedrock
topography, and this was corrected for using Newtonian gravity.  The correction from this is actually the cause of the
non-linearity of the data points in Fig.\ref{fig:Greenland}.  A complete analysis would require that the effect of
this rock terrain be also computed using the new theory of gravity, but this was not done.    
Using $G_N=6.6742\times10^{-11}$ m$^3$s$^{-2}$kg$^{-1}$, which is the current CODATA value, see
sect.\ref{sect:Gdata}, we obtain from a least-squares fit of   the linear term in (\ref{eqn:E15}) to the data
points in Fig.\ref{fig:Greenland} that $\alpha^{-1}=137.9\pm 5$, which equals the value of the fine structure
constant 
$\alpha^{-1}=137.036$ to within the errors, and for this reason we identify the  constant $\alpha$ in (\ref{eqn:E1})
as being the fine structure constant. 
 The first  analysis \cite{alpha,DMtrends} of the Greenland Ice Shelf data incorrectly  assumed that the ice density was
930 kg/m$^3$ which gave $\alpha^{-1}=139\pm 5$. However trapped air reduces the standard ice density to  the ice
shelf density of 920 kg/m$^3$, which brings the value of $\alpha$ immediately into better agreement with the
value of $\alpha=e^2/\hbar c$  known from quantum theory.

\section{Nevada Test Site Borehole Data}

\begin{figure}[t]
\hspace{45mm}\includegraphics[scale=0.27]{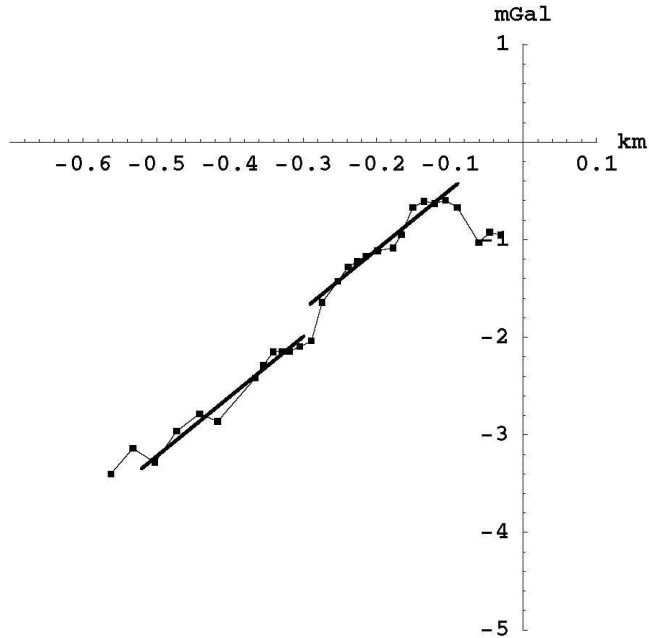}
\caption{\small{  The data shows the gravity residuals for the Nevada U20AL borehole  Airy measurements of the
$g(r)$  profile \cite{Thomas90},  defined as
$\Delta g(r) = g_{Newton}-g_{observed}$, and measured in mGal, plotted against depth in km.
This residual shows regions of linearity interspersed with regions of non-linearity, presumably arising from layers with
a density different from the main density of 2000 kg/m$^3$. Density changes generate a change in the (arbitrary)
residual offset.  From a least-squares simultaneous fit of the linear form in (\ref{eqn:E15}) to the slope of the
four linear regions  in this data  and that in Fig.\ref{fig:Nevada1} for the data from borehole U20AK in
Fig.{\ref{fig:Nevada1}, we obtain $\alpha^{-1}=136.8\pm 3$. The two fitted regions of data are shown by the two straight
lines here and in  Fig.{\ref{fig:Nevada1}}. }}  
\label{fig:Nevada2}}\end{figure}

\begin{figure}[t]
\hspace{45mm}\includegraphics[scale=0.27]{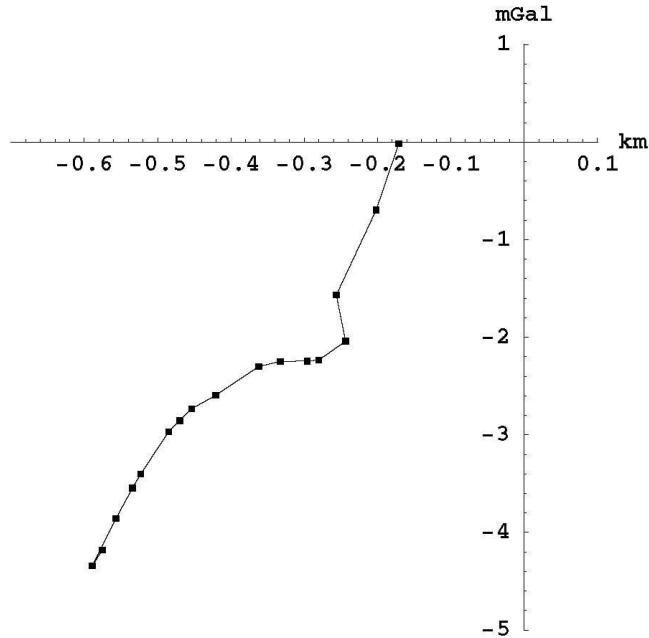}
\caption{\small{ The data shows the gravity residuals from the third Nevada U20AO borehole  Airy measurements of the
$g(r)$  profile \cite{Thomas90}}. This data is not  of sufficient linearity, presumably due to non-uniformity of
density, to permit a fit to the linear form in (\ref{eqn:E15}), but is included here for completeness. There is an
arbitrary offset in the residual.
\label{fig:Nevada3}}\end{figure}

Thomas and Vogel \cite{Thomas90} performed another borehole experiment at the Nevada Test Site in 1989 in which they
measured the gravity gradient as a function of depth, the local average matter density, and the above ground
gradient, also known as the free-air gradient.  Their intention was to test the extracted $G_{local}$ and compare
with other values of
$G_N$, but of course using the Newtonian theory. The Nevada boreholes, with typically 3 m diameter, were drilled as a
part of the U.S. Government tests of its nuclear weapons. The density of the rock is measured with a $\gamma-\gamma$
logging tool, which is essentially a $\gamma$-ray attenuation measurement, while in some holes the rock density was
measured with a coreing tool. The rock density was found to be  2000 kg/m$^3$, and is dry. This is the density used
in the analysis herein.   The topography for 1 to 2 km beneath the surface  is dominated by a series of overlapping
horizontal lava flows and alluvial layers. Gravity residuals from three of the bore holes are shown in  
Figs.\ref{fig:Nevada1}, \ref{fig:Nevada2} and \ref{fig:Nevada3}. All gravity measurements were corrected for the
earth's tide, the terrain on the surface out to 168 km distance, and the evacuation of the holes. The gravity
residuals arise after allowing for, using Newtonian theory, the local lateral mass anomalies but assumed that the matter
beneath the holes occurs in homogeneous ellipsoidal layers.   Here we now report a detailed analysis of the
Nevada data.  First we note that the gravity residuals from borehole  U20AO, Fig.\ref{fig:Nevada3}, are not
sufficiently linear to be useful. This presumably arises from density variations caused by the layering effect. 
For boreholes UA20AK, Fig.\ref{fig:Nevada1}, and UA20AL, Fig.\ref{fig:Nevada2}, we see segments where the gravity
residuals are linear with depth, where the density is the average value of 2000 kg/m$^3$, but interspersed by layers
where the residuals show non-linear changes with depth. It is assumed here that these non-linear regions are caused
by variable density layers.  So in  analysing this data we have only used the linear regions, and a simultaneous
least-squares fit of the slope of (\ref{eqn:E15}) to the slopes  of these four linear regions gives
$\alpha^{-1}=136.8\pm3$, which again is in extraordinary agreement with the value of $137.04$ from quantum
theory.   Here  we again used $G_N=6.6742\times10^{-11}$ m$^3$s$^{-2}$kg$^{-1}$, as for the
Greenland data analysis,

\section{Ocean   Measurements }
The ideal Airy experiment would be one using the ocean, as all relevant physical aspects are accessible. Such 
an experiment was carried out by Zumberge {\it et al.} in 1991 \cite{Zumberge1991} using submersibles. Corrections
for sea floor topography, seismic profiles and  sea surface undulations
 were carried out.  However a true Airy experiment appears not to have been performed. That would have required the
measurement  of the above and below sea-surface gravity gradients. Rather only the below sea-surface
gradients were measured, and compared with a predicted gravity gradient using  the
density of the water and a laboratory value of $G_N$ from only one such experiment and, as shown in Fig.\ref{fig:GData},
these have a large uncertainty.  Hence this experiment does not permit an analysis of the data of the form applied to
the Greenland and Nevada observations. The value of $G_N$ from this ocean experiment is shown in   Fig.\ref{fig:GData}
as experiment \#12.
  
\section{$G$ Experiments \label{sect:Gdata}}

\begin{figure}
\hspace{45mm}\includegraphics[scale=0.25]{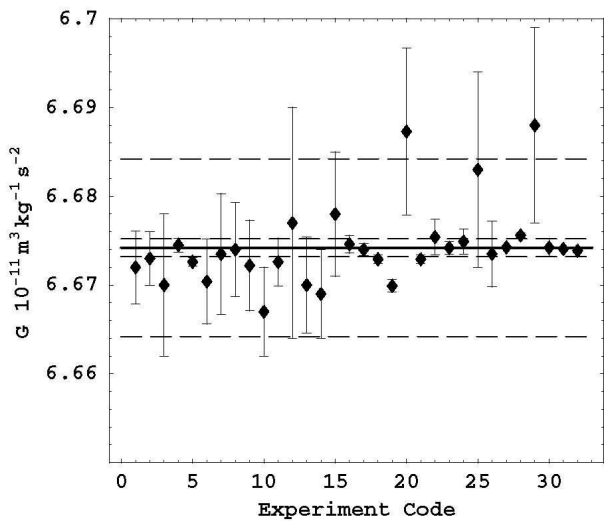}
\caption{\small{Results of precision measurements of $G_N$ published in the 
last sixty years in which the Newtonian theory was used to analyse the data.  These results 
show  the presence of a  systematic effect, not in the Newtonian theory, of fractional size up to $\Delta G_N/G_N \approx
\alpha/4$, which corresponded with the 1998 error bars on $G_N$ (outer dashed  lines), with the full line being the
current CODATA value of $G_N=(6.6742(10)\times 10^{-11} $ m$^2$s$^{-2}$kg$^{-1}$. In 2005 CODATA
\cite{CODATA2002} reduced the error bars by a factor of 10 (inner dashed lines) on the basis of  some recent
experiments, and so neglecting the presence of the systematic effect.   }   
\label{fig:GData}}\end{figure}

The new theory of gravity, given in (\ref{eqn:E1}) for the case of zero vorticity and in the non-relativistic limit, 
is a two-parameter theory; $G$ and $\alpha$.  Hence in experiments to determine $G$ (or $G_N$) we expect to see
systematic discrepancies if the Newtonian theory is used to analyse the data.   This is clearly the case as shown in 
 Fig.\ref{fig:GData} which shows the results of such analyses over the last 60 years. The fundamental problem is that
non-Newtonian effects of size approximately  $\Delta G_N/G_N \approx \alpha/4$ are clearly evident, and effects of this
size are expected from  (\ref{eqn:E1}).  To correctly analyse data from these experiments the full theory in 
(\ref{eqn:E1}) must be used, and this would involve (i) computing the velocity field for each configuration of the test
masses, and then (ii) computing the forces by using  (\ref{eqn:E3}) to compute the acceleration field. These
computations are far from simple, especially when the complicated matter geometries of recent experiments need to be
used.  Essentially the flow of space results in a non-Newtonian effective `dark matter' density in (\ref{eqn:E5}).
This results in deviations from Newtonian gravity which are of order $\alpha/4$.  The prediction is that when
laboratory Cavendish-type experiments are correctly analysed the data will permit the determination of both $G_N$ and
$\alpha$, and the large uncertainties in the determination of $G_N$ will no longer occur.  Until then the value of
$G_N$ will continue to be the least accurately known of all the fundamental constants.   Despite this emerging
insight CODATA\footnote{CODATA is the Task Group on Fundamental Constants of the
Committee on Data for Science and Technology, established in 1969. } in 2005
\cite{CODATA2002} reduced the apparent uncertainties in $G_N$ by a factor of 10,
and so ignoring the manifest presence of a systematic effect.  The occurrence of the fine structure constant
$\alpha$, in giving the magnitude of the spatial self-interaction effect in (\ref{eqn:E1}), is a fundamental
development in our understanding of 3-space and the phenomenon of gravity. Indeed the implication is that
$\alpha$ arises here as a manifestation of quantum processes inherent in 3-space.

\section{Some History}
Here we have simply applied the new two-parameter theory of 3-space, and hence of  gravity, to the existing
data from borehole experiments.  However the history of these experiments shows that, of course, the nature
of the gravitational anomaly had not been understood, and so the  implications for fundamental physics that
are now evident could not have been made. The first indications that some non-Newtonian effect was being
observed arose from Yellin \cite{Yelin1968} and Hinze {\it et al.} \cite{Hinze1978}. It was Stacey {\it et
al.}  in 1981 \cite{Stacey1981b,Stacey1981,Holding1986} who undertook systematic studies at the Mt.Isa Mine
in Queensland, Australia.  In the end a mine site is very unsuited for such a gravitational anomaly
experiment as by their very nature mines have non-uniform poorly-known density and usually, as well, 
irregular surface topography. In the end it was acknowledged that the Mt.Isa Mine data was unreliable.
Nevertheless those reports motivated the Greenland, Nevada and Ocean experiments, as well as above-ground
tower experiments \cite{Thomas1989}, all with the assumption that the non-Newtonian effects were being caused
by a modification to Newton's {\it inverse square law} by an additional short-range force - which also
involved the notion of  a possible``5th-force" \cite{Fischbach}.  However these interpretations were not
supported by the data, and eventually the whole phenomenon of these gravitational borehole anomalies was forgotten.

\section{Conclusions\label{sect:conclusions}}
We have extended the results from an earlier analysis  \cite{alpha,DMtrends} of the Greenland Ice Shelf
borehole $g$ anomaly data by  correcting the density of ice from the assumed value to the actual value. This
brought the extracted value of $\alpha$ from approximately 1/139 to approximately 1/137, and so into even
closer agreement with the quantum theory value. As well the analysis was extended to the Nevada borehole
anomaly data, again giving $\alpha\approx 1/137$. This is significant as the rock density is more than
twice the ice density. As well we have included the previous results \cite{BH} from analysis of the
blackhole masses  in globular clusters and elliptical ``spherical" galaxies, which gave $\alpha \approx
1/134$, but with larger uncertainty. So the conclusion  that $\alpha$ is actually the fine structure constant from
quantum theory is now extremely strong. These results, together with the successful explanation for the so-called
spiral galaxy ``dark-matter" effect afforded by the new theory of gravity, implies that the Newtonian theory of
gravity
\cite{Newton} is fundamentally flawed, even at the non-relativistic level, and that the disagreement with
experiment and observation can be of fractional order $\alpha$, or in the case of spiral galaxies and blackholes,
extremely large.  This failure implies that General Relativity, which reduces to the Newtonian theory in the
non-relativistic limit, must also be considered as flawed and disproven.  

This research is supported by an Australian Research Council Discovery Grant.

\end{document}